# Reconfiguring health services to reduce the workload of caregivers during the COVID-19 outbreak using an open-source scalable platform for remote digital monitoring and coordination of care in hospital Command Centres

Philippe Ravaud[1,2], Franck le Ouay[3*], Etienne Depaulis[3], Alexandre Huckert[3], Bruno Vegreville[4], Viet-Thi Tran[1,2]

[1] University of Paris, CRESS, INSERM, INRA, F-75004 Paris, France

[2] Assistance Publique Hôpitaux de Paris, F-75004 Paris France

[3] Lifen, Paris France

[4] Inato, Paris, France

**Words:** 1167

**Corresponding author:**

Frank le Ouay
Lifen
franck@lifen.fr

**Contributorship statement:**

PR, FLO, ED, AH, BV and V-TT designed the remote monitoring platform and the Command Centres. PR and V-TT drafted the online questionnaires and the decision rules, with the help of clinicians. FLO, ED, AH and BV developed the software. V-TT drafted the manuscript. All authors provided critics and revised on the manuscript. All authors approved the manuscript.

**Dissemination to patients' organizations is not applicable.**

**There was no public and patient involvement in the conception of this manuscript**

## Abstract (82 words)

The Covid-19 outbreak threatens to saturate healthcare systems in most Western countries. We describe how digital technologies may be used to automatically and remotely monitor patients at home. Patients answer simple self-reported questionnaires and their data is transmitted, in real time, to a Command Centre in the nearest reference hospital. Patient reported data are automatically filtered by algorithms to identify those with early warning signs. Open-source code of all software components required to deploy the remote digital monitoring platform and Command Centres is available.

**Context**

The coronavirus disease 2019 (Covid-19) has been declared a public health emergency of international concern[1]. On the 11th of March 2020, 118 162 cases were confirmed in 113 countries, with 4 290 deaths [2]. The Covid-19 outbreak threatens to saturate healthcare systems in Western countries which are already at breaking point dealing with routine demands[3]. Indeed, the virus transmits easily (R0 between 1.4 and 3.28) and its current management involves mainly hospital care, despite the fact that most patients who will contract the virus will have few symptoms and long term effects [3-5]. Naïve calculations show that this strategy will not be sustainable as the epidemic progresses [6]. To anticipate for the escalation of the epidemic, since the 7th of March 2020, patients with confirmed Covid-19 without signs of severity in France can be managed at home by their general practitioner [7]. Similarly, in the United Kingdom, a 24 hour, seven day a week, service has been set-up to manage patients who "do not require immediate admission" to the hospital", at home [5]. Yet, these solutions rely heavily on in-person management of patients and will still place an important workload on general practitioners and other care professionals involved.

Our proposal is to exploit digital technologies to remotely monitor patients at home. Data from remote monitoring of patients are transmitted to a Command Centre, in the nearest reference hospital, and automated algorithms triage patients with early warning signs. This will spare human time and let physicians focus on patients who may need specific care actions (**Figure, Supplementary Material 1**). In addition to reducing contacts with care professionals and risks of contamination, remote monitoring of patients and automated decisions may alleviate the workload of caregivers and delay the expected disorganization of care structures.

In a few days, public health researchers and clinicians from the Assistance Publique Hôpitaux de Paris and University of Paris, software architects and developers co-constructed a prototype for the remote monitoring platform (**Supplementary Material 2**). Open-source code of all software components required to deploy the remote monitoring platform and Command Centres was developed by Lifen and is available at: https://github.com/lifen-labs/covid.

## Remote monitoring for patients in quarantine

Patients with confirmed Covid-19 are assessed by clinicians (in hospital or in community) for: 1) absence of initial signs of severity (based on their age, comorbidities, initial presentation of the disease); 2) their ability to be quarantined at home (e.g., absence of a psychiatric disorder or of a loss of autonomy); and 3) their ability to perform the remote monitoring at home (e.g., basic computer literacy, smartphone availability). If all criteria are fulfilled and if the patient consents for remote digital monitoring, he is sent home with instructions for quarantine [8]. Information is automatically sent to the patient's general practitioner, informing them that one of their patients has been confirmed with Covid-19 and is now being monitored at home.

Remote follow-up of patients at home was designed to be minimally disruptive. It consists of a self-reported questionnaire, once or twice a day. Patients receive a text-message with a direct secure link to an online questionnaire. No login is required. Questionnaires involve <10 items and collect self-reported symptoms with validated tools (e.g., temperature, dyspnoea, pain) and quarantine information (e.g., psychological state regarding the quarantine and the disease, change of the people who are at home with them). In case of emergency, patients can contact the Command Centre or the National emergency number.

## Command Centres

Command Centres are located in hospitals and involve human personnel, including physicians and nurses who will analyse the constant influx of information from the remote monitoring. These people are equipped with real-time and decision-support tools, and assess whether patient care needs to be modified. They take necessary actions (e.g., intensifying monitoring, sending medical assistance, calling the patient for reassurance etc.) or provide feedback to patients and General Practitioners.

Each time a patient sends new information by completing a self-reported questionnaire, their data are updated in real time. Automatic algorithms flag patients, using pre-defined decision rules, in four categories:

- "Green" patients are asymptomatic and have no problems with the quarantine. No action from clinicians is required for these patients. Automatic messages are sent to reassure patients and to remind them to continue completing the regular questionnaires.
- "Yellow" patients are stable with no signs of severity. No action from clinicians is required for these patients. Automatic messages are sent to reassure patients and to remind them to continue completing the regular questionnaires.
- "Orange" patients are those with a recent change of symptoms and who may require closer monitoring. When a patient is flagged "Orange", the frequency of questionnaires is increased. In addition, an action is required from clinicians.
- "Red" patients are those with rapid evolution of symptoms, signs of severity and/or those who have problems with the quarantine. Rapid action is required from clinicians.
- "Patients who did not complete the questionnaire after 8h". Patients are highlighted and are called by the Command Centre.

Whenever new data is received at the Command Centre, a summary of the patient status and of decisions taken is sent to their GP.

To manage patients, the human personnel in Command Centres have access to dashboards where they can 1) visualize all patients enrolled in the remote monitoring, their main symptoms, and their flags; 2) examine a given patient's data and take actions; or 3) zoom out to examine the state of the Command Centre regarding specific subgroups of patients (e.g., those requiring actions, severe patients, etc.). A video demonstration of the Platform is available **(Supplementary Material 3)**.

## Scalability and Future evolutions

The platform was designed to be compatible with nationwide deployment, in various size hospitals.

The remote digital monitoring platform was also envisioned to evolve according to the evolution of the epidemic. First, the platform could be modified for other purposes, such as estimating the number of patients developing the disease by using simple self-reported questionnaires to identify the moment when asymptomatic contact subjects develop the disease. Secondly, the platform was thought to be compatible with the potential enrichment of patient-reported information with data from biometric monitoring devices, such as pulse oximeters (these could be either provided to patients or via patients' own smartphones as in a Bring Your Own Device approach[9]). This could help further pinpoint patients requiring hospital care and spare precious hospital beds. Thirdly, our platform could be integrated in a larger ecosystem of Covid-19 management integrating the remote follow-up of patients using different communication channels (e.g., telephone, smartphone application, etc.) according to the age, ability of the patient to use a smartphone, or severity of the disease etc. Finally, the remote digital monitoring of patients will provide continuous streams of data which will automatically update interactive dashboards about the local or regional state of the epidemic and about the capacity of the different Command Centres. In addition, data from all command centres can be pooled to provide real-time data visualizations on the state of the epidemic at national level or to constitute a research database.

## Acknowledgements

We thank greatly Prof Xavier Lescure (Infectious Diseases), Prof Enrique Casalino (Emergency Medicine) and Dr François Grolleau (Intensive Care Medicine) who helped us build the clinical questionnaires and decision rules. We also thank Dr Youri Yordanov (Emergency Medicine) for useful discussions about this paper. We thank Elise Diard for her help in drafting figures.

## Conflicts of Interest

Franck le Ouay, Etienne Depaulis and Alexandre Huckert are co-founders of Lifen. Bruno Vegreville is employed by Inato. Philippe Ravaud holds shares in Inato.

# Figures

Figure. Key elements of the remote digital monitoring platform and Command Centres

# Supplementary Material

Supplementary Material 1. Architecture diagram of the remote digital monitoring platform and Command Centres

Supplementary Material 2. Development steps of the remote digital monitoring platform and Command Centres

Supplementary Material 3. Video demonstration of the remote digital monitoring platform and Command Centres

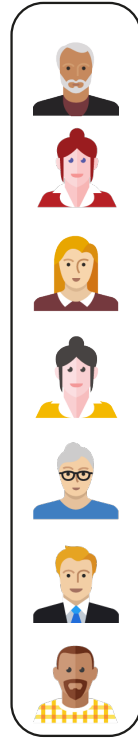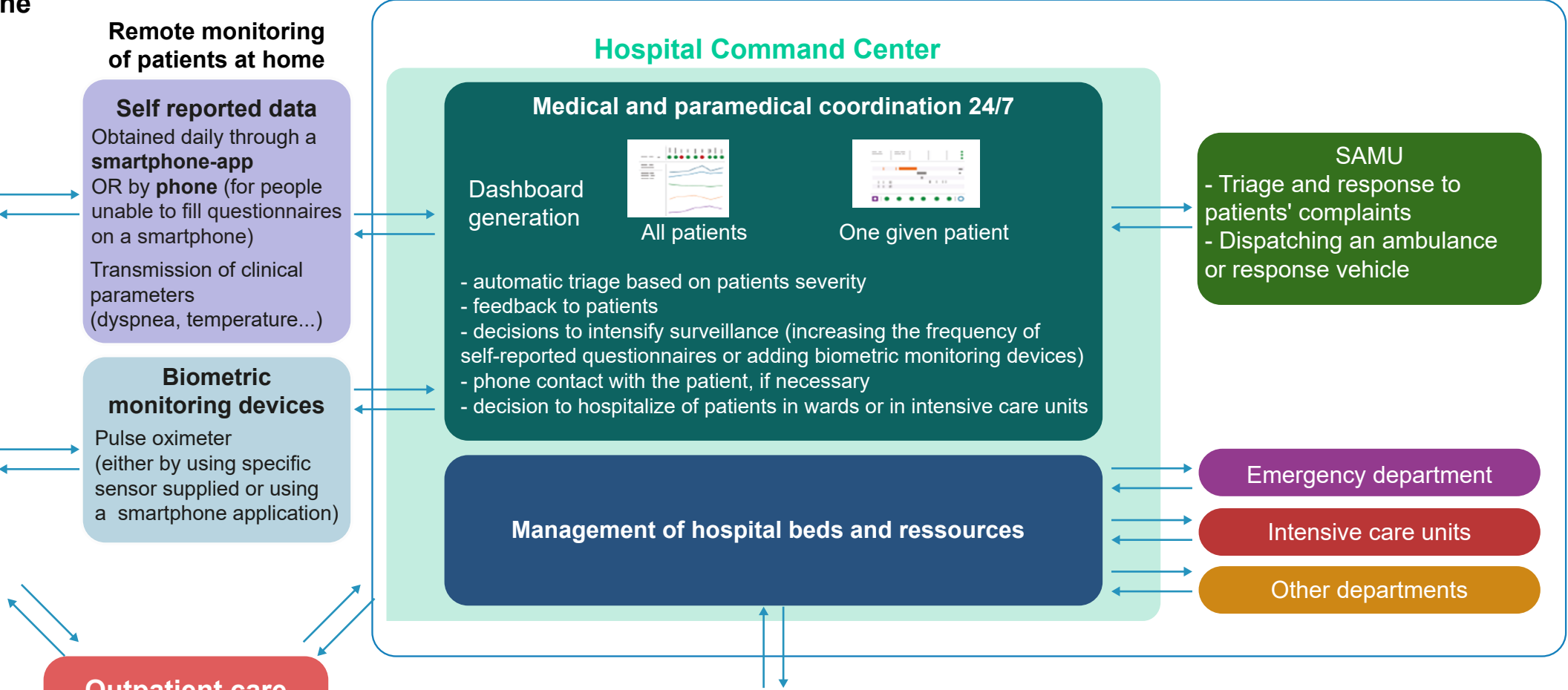

# Supplementary Materials

## Supplementary Material 1. Architecture diagram of the remote digital monitoring platform and Command Center

**Flowchart for a patient**

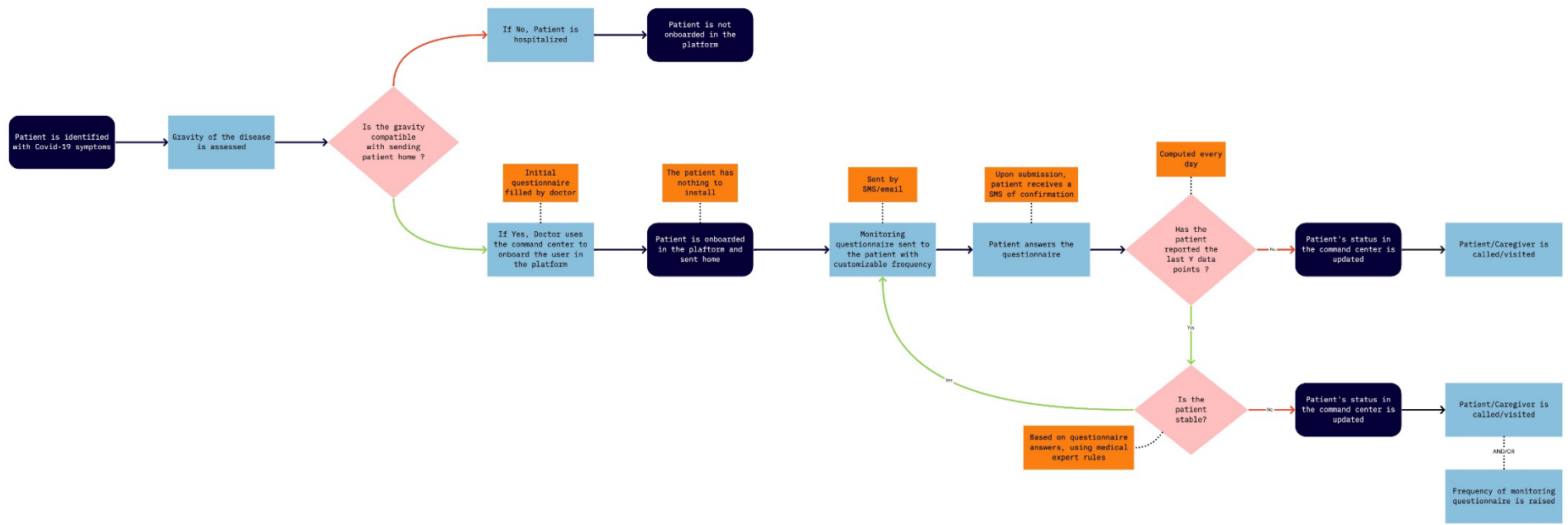

**Supplementary Material 2. Development steps of the remote digital monitoring platform and Command Center**

## DAY 1
*Monday March 2*

**Back-office available**

- **Management & access rights** (administrators, doctors)

- Possibility to **enroll patients**

- Form to **fill up patient informations** when initiating a remote monitoring. This form is to be used by HCPs

**Remote-monitoring form available**

- Mobile-first web **responsive interface**

- Creation of a system to **generate secret and secure links** (valid for 24h) for each patient and each form sent

## DAY 2
*Tuesday March 3*

- Final updates on form content based on the recommendations of medical experts

- Set up periodic text-messaging service
  *We are using Twilio for sending text messages.*

- Adding **an autoresponder voice and text** to guide patients (in case they respond to the text message or call the number)

- Setup a dashboard to **analyse results and create reports**
  *We simply plugged Metabase, because we know it well, for using it everytime at Lifen.*

## DAY 3
*Wednesday March 4*

- We decided to work from Paris's central hospital (Hotel Dieu) **to be able to iterate very quickly on the produdct** with clinicians and the epidemologic team

- Added **Patient statut computation** to categorize them in 4 categories of severity (green, yellow, orange, red), to be ajusted based on feedback from the field

- Dashboard to **manage and visualize patients** per category
  *Adding filters, search, and real-time updates*

- Added instructions to **help patients understand** how to measure themselves correctly

- Complete validation of platform with **medical experts**

**Supplementary Material 3. Video demonstration of the remote digital monitoring platform and Command Center**

https://youtu.be/wwDJNR6SKmI

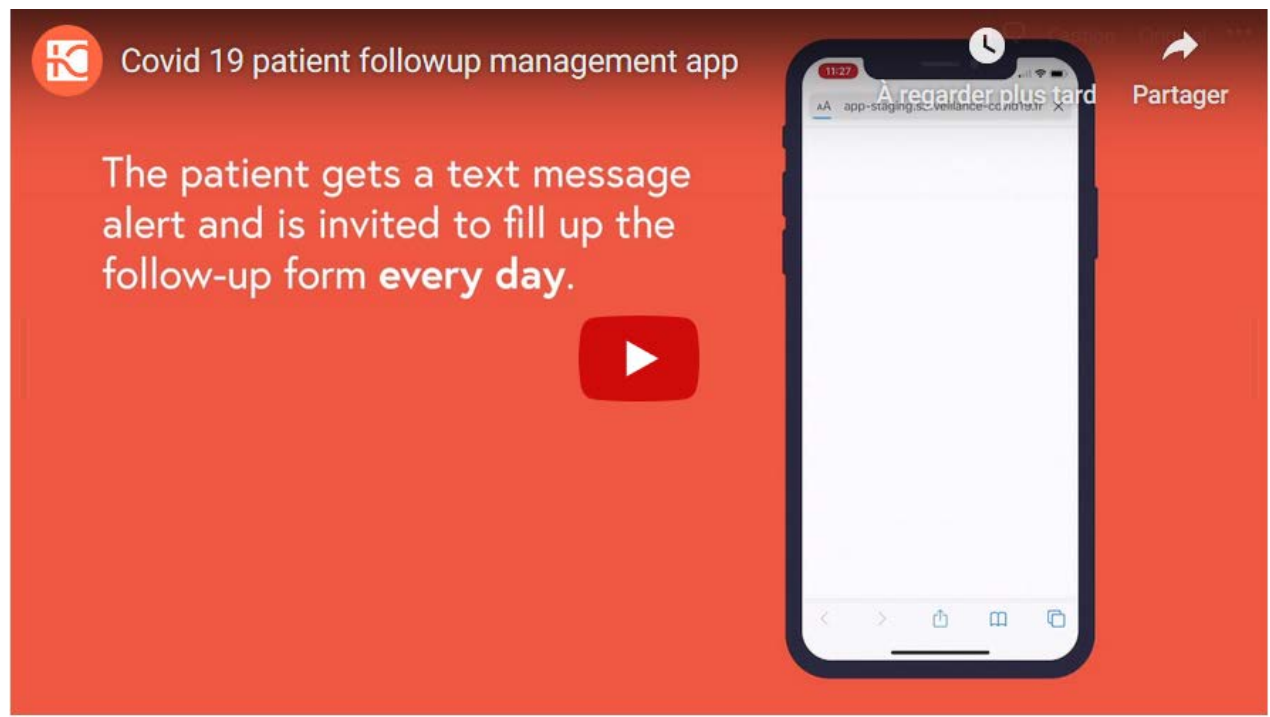